\documentclass{desyproc}

\begin{document}
\title{Stellar Evolution Bounds on the ALP-Photon Coupling: new Results and Perspectives 	}

\author{{\slshape Maurizio Giannotti}\\[1ex]
Barry University, Miami Shores, US}

\contribID{familyname\_firstname}

\confID{300768}  
\desyproc{DESY-PROC-2014-03}
\acronym{Patras 2014} 
\doi  

\maketitle

\begin{abstract}
Stellar evolution considerations are of fundamental importance in our understanding of the axion/ALP-photon coupling, $ g_{a\gamma} $. Helium burning stars are the best laboratories to study this coupling. Here, we will review the bounds from massive and low mass helium burning stars, and present a new analysis of the bound from the horizontal branch stars. The analysis provides the strongest bound to date on $ g_{a\gamma} $ in a wide mass range.
\end{abstract}

\section{Introduction}

For several decades stellar evolutionary arguments have provided an invaluable tool to understand various properties of \emph{light, weakly interacting} particles, offering an alternative to terrestrial experiments and often providing even stronger bounds over wide regions of the parameter space.
%
%
%
Some of the most successful examples 
include the study of nonstandard neutrino properties~\cite{Bernstein:1963qh,Dicus:1976ra,Heger:2008er,Viaux:2013lha},
majorons \cite{Georgi:1981pg}, novel baryonic or leptonic forces \cite{Grifols:1986fc}, unparticles \cite{Hannestad:2007ys}, extra-dimensional photons \cite{Friedland:2007yj}, axions~\cite{Raffelt:1985nk,Raffelt:1987yu,Friedland:2012hj,Viaux:2013lha,Ayala:2014pea}
and, in general, WISPs (Weakly Interacting Slim Particles)~\cite{Baker:2013zta}.

Here, we are interested in the axions, and in particular in its coupling to photons
\begin{equation}\label{Eq:agg}
	 L=-\frac{g_{a\gamma}}{4}a F\tilde F = g_{a\gamma} \textbf{E}\cdot \textbf{B}\,.
\end{equation}
Standard (QCD) axions are expected to satisfy a simple relation between mass and coupling,  $(m_{a}/1\mbox{ eV})= 0.5\, \xi \, g_{10}$, 
where $ g_{10}=g_{a\gamma}/(10^{-10}\rm{GeV}^{-1})$ and $\xi$ is of order~1 in many motivated axion models. 
This defines the so called \emph{axion-line} in the $ m_a-g_{a\gamma} $  plane.
However, models of QCD-axions which do not satisfy this relation existed for a long time~\cite{Rubakov:1997vp,Berezhiani:2000gh,Gianfagna:2004je}.
In addition, recently a considerable attention has been devoted to the so-called Axion-Like-Particles (ALPs), light pseudoscalr particles, coupled to photons as in Eq. (\ref{Eq:agg}) but not necessarily on the axion-line. 
Such particles emerge naturally in various extensions of the Standard Model (for a recent review see~\cite{Baker:2013zta}) and are phenomenologically motivated by a series of unexplained astrophysical observations (see, e.g., \cite{Carosi:2013rla} and references therein). 

Several experiments are currently involved in the axion/ALP search.
In particular, the CERN Axion Solar Telescope (CAST)~\cite{Andriamonje:2007ew}, a $3^{\rm th}$ generation axion helioscope based at CERN, provides the strongest terrestrial bound, $ g_{10}\lesssim 0.88 $, on light ALPs ($ m_a<$ a few eV).
A larger and more sensitive ($4^{\rm th}$ generation) helioscope, the International Axion Observatory (IAXO,~\cite{Irastorza:2011gs}), recently proposed, would allow the exploration of the parameter region about an order of magnitude below the current CAST bound on $ g_{a\gamma} $ in a wider mass range. 



However, the strongest current bounds on the axion photon coupling are derived from astrophysical considerations.
%
In particular, the analysis of the evolution of intermediate mass stars, $ M =8-12\,M_\odot$, leads to the bound $ g_{10}\leq 0.8 $, while the analysis of globular cluster stars provides the constraint $ g_{10}\leq 0.66 $. 
Both bounds apply to a wide mass range, up to a few 10 keV. 


\section{Helium burning stars and axions}

Axions or ALPs with mass below a few keV could be produced in stellar interiors via the Primakoff process -- 
the conversion of a photon into an axion  in the electric field of nuclei and electrons in the stellar plasma~\cite{Raffelt:1985nk}.
%
The Primakoff process is particularly efficient in the core of He burning stars, at a temperature of about $ 10^8 $K and density of $ 10^3 - 10^4 $ g$ \cdot $cm$ ^{-3} $.
At higher temperatures, though the axions production increases, the pair neutrino process starts dominating and, by the time Carbon burning processes start, it becomes the main cooling mechanism of the star. 

Stars of low and high mass show qualitative differences during the core He-burning phase and it is interesting to study them separately.

\subsection{Massive stars: the Cepheids bound}
Stars a few times more massive than the sun go through the so called \emph{blue loop} phase while burning helium.
This stage starts with a contraction during which the star becomes hotter (bluer), followed by a new expansion and cooling which brings the star back to the red region to the right of the HR diagram. 

Interestingly, the beginning of the blue loop is fairly independent from the details of the core burning and it is fairly unaffected by axion emission, at least in the range of coupling of interest to us. 
However, the end of the loop depends on the amount of helium in the core.
The presence of an efficient cooling mechanism speeds up the helium consumption and so can induce the loop to terminate early or even prevent it from starting. 
Though known for several decades (see, e.g., \cite{Lauterborn:1971nva}), this observation was applied to the axion case only recently~\cite{Friedland:2012hj} and it was shown that a value of $  g_{10}\gtrsim 0.8 $ would eliminate the blue loop evolution for all stars in the mass range $ M \sim 8-12\,M_\odot$.

It is particularly interesting to note that the blue loop extends on the instability strip and so it is necessary to explain the existence of Cepheid stars, whose properties are well understood and very well measured. 
The absence of Cepheids of mass $ M \sim 8-12\,M_\odot$ would imply a gap in the observed oscillation periods (mass and period of a Cepheid are related), a fact which is not observed~\cite{Friedland:2012hj}.

The relation of this bound to the Cepheid stars makes it quite robust from an observational point of view, since the periods of these stars are very well measured. 
However, the numerical modeling does present various uncertainties, especially in relation to the convection prescription. 
Notably, some amount of overshooting would reduce the extension of the loop and eventually eliminate it. 
The perfect convection prescription is, unfortunately, not known. 
However, there are indications in favor a non-minimal overshooting (see, e.g., \cite{Keller:2008yu}).
If so, the Cepheid bound on the axion photon coupling could be somewhat lowered. 
A full analysis of the uncertainties in this bound is, at the moment, missing.

\subsection{Low mass stars: the HB bound}
Low mass globular cluster (GC) stars are also very efficient laboratories to study the axion-photon coupling.
\begin{wrapfigure}{R}{0.5\textwidth}
   \centerline{\includegraphics[width=0.45\textwidth]{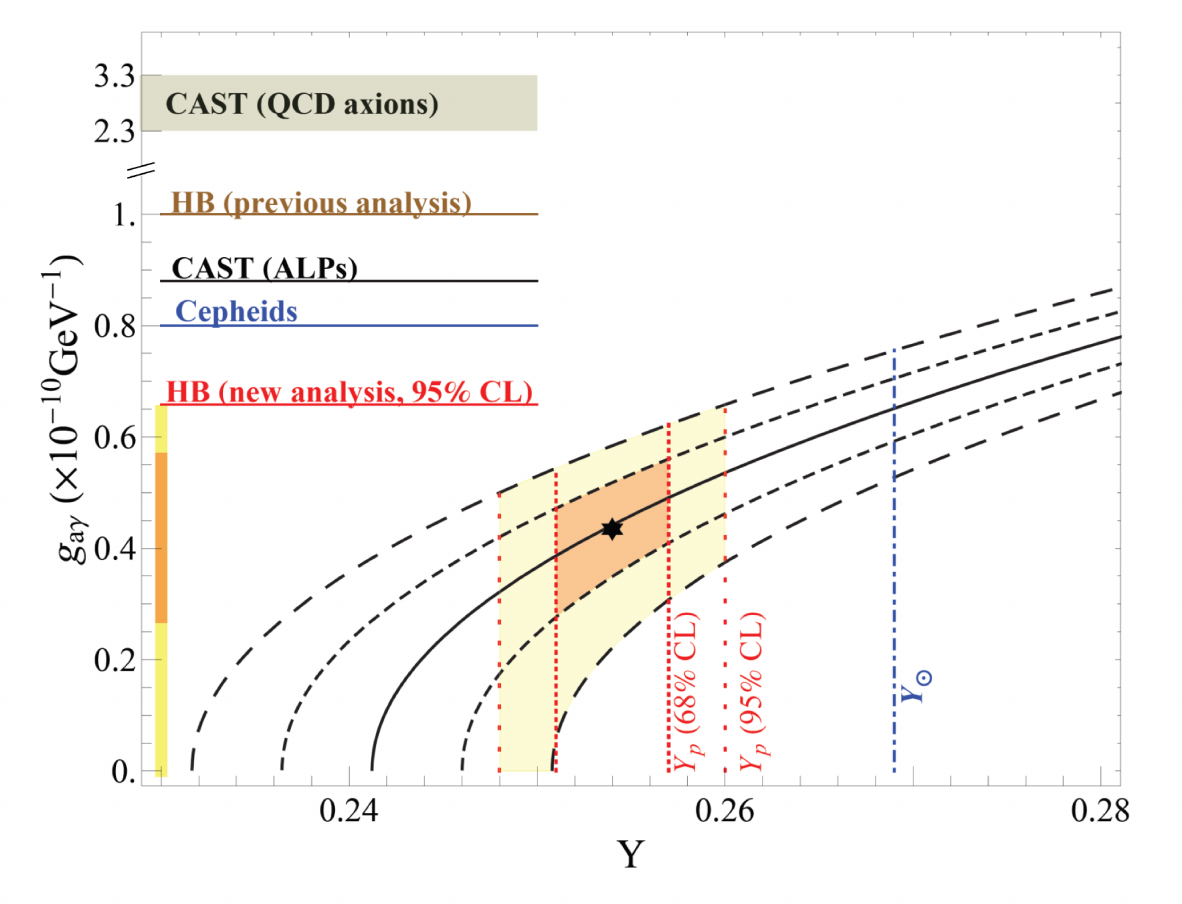}}
\caption{Constraints on $g_{a\gamma}$.}\label{fig:G10VsY}
\end{wrapfigure}
GCs, gravitationally bound systems of stars populating the Galactic Halo, are among the oldest objects in the Milky Way. 
Hence only low mass stars (M$\lesssim 0.85$~M$_\odot$) are still alive and, therefore, observable. 
A typical CG harbors a few millions  stars, so that the various evolutionary phases are well populated and distinguished from each other. 
It was recognized early on that axions coupled to photons would significantly reduce the lifetime of stars in the 
horizontal branch (HB) evolutionary stage, corresponding to the 
core He burning phase, while producing negligible changes during the preceding red giant branch (RGB) evolution~\cite{Raffelt:1987yu}.
So, $g_{a\gamma}$ can be constrained by measurements of the $ R $ parameter,  
$R= {N_{\rm HB}}/{N_{\rm RGB}}$, 
which compares the numbers of stars in the HB  ($N_{\rm HB}$) and in the upper portion of the RGB ($N_{\rm RGB}$). 

The original analyses of the axion bound from HB stars
were based on the assumption that the measured $R$ parameter is well
reproduced, within 30\%, by models of GC stars without including axion cooling.
This approach, however, neglects the effects of the initial helium mass fraction Y which also affects the value of the $ R $ parameter.
Quantitatively we found~\cite{Ayala:2014pea},
\begin{eqnarray}\label{eq:erreth}
R (g_{a \gamma}, Y)=6.26\, Y-0.41\, g_{10}^2 -0.12\,.\nonumber
\end{eqnarray}
According to the above relation, even a considerable decrease of the HB lifetime caused by a large value of $g_{a\gamma}$ could be compensated by a suitable increase in the assumed He content.
Because of this evident degeneracy, a proper evaluation of the axion constraint from the $R$ parameter relies on our 
knowledge of the He abundance in the GCs.

This effect has been taken into account in our recent work~\cite{Ayala:2014pea}.
In this new analysis, the recent data for the R parameter~\cite{Salaris:2004xd} have been compared with measurements of the helium mass fraction Y in GCs~\cite{Izotov:2013waa,Aver:2013wba}. 
The results are shown in Fig.~\ref{fig:G10VsY}.
The vertical lines indicate the 
$1\sigma$ and $2\sigma$ regions of Y and the dot-dashed vertical line the
He abundance in the early solar system Y$_{\odot}$, provided here as a  reference value (see discussion in \cite{Ayala:2014pea}). 
The bent curves show $R_{\rm ave}$ and its 1$\sigma$ and the 2$\sigma$ ranges.
The  shaded area  delimits the combined 68\% CL (dark) and 95\% CL (light) for Y and $R$.
The vertical rectangles indicate the 68\% CL (dark) and 95\% CL (light) for $g_{a\gamma}$.
Previous bounds are also shown.
		
The resulting constrain $ g_{10}\leq 0.66 $ (at 95\% CL) represents the strongest limit on $g_{a\gamma}$ for QCD axions and ALPs in a wide mass range.

\section{Discussion and conclusion}

Helium burning stars are an excellent laboratory to study the ALP-photon coupling and provide the strongest bounds in a wide mass range which extends up to a few 10 keV.

The analysis from massive stars, which provides the bound $ g_{10}\leq 0.8$, shows an interesting connection between Cepheids and particle physics and could possibly be applied to other scenarios in the low energy frontier.

The recent analysis of globular clusters stars provides an even stronger bound, $ g_{10}\leq 0.66$ at 95\% CL, and, for the first time, an analysis of the uncertainties. 

From the Figure~\ref{fig:G10VsY}, it appears that a weakly coupled axion or axion-like particle could improve the relation between the observed $ R $ parameter and the helium mass fraction. 
However, the statistical significance of this result is very low and we prefer, for the moment, to use this analysis only to extract the upper bound on the coupling. 
Additional investigation may reveal if the effect is just due to poor statistics or indicates new physics.
In any case, it is intriguing that the interesting parameter region is well in reach of the next generation experiments, notably ALPs II~\cite{Bahre:2013ywa} and IAXO~\cite{Irastorza:2011gs}.

\section{Bibliography}


\begin{footnotesize}

\end{footnotesize}



\begin{thebibliography}{99}
%
\bibitem{Bernstein:1963qh}
  J.~Bernstein, M.~Ruderman and G.~Feinberg,
  Phys.\ Rev.\  {\bf 132} (1963) 1227.
  
  
\bibitem{Dicus:1976ra}
  D.~A.~Dicus and E.~W.~Kolb,
  Phys.\ Rev.\ D {\bf 15} (1977) 977.

\bibitem{Heger:2008er}
  A.~Heger, A.~Friedland, M.~Giannotti and V.~Cirigliano,
  Astrophys.\ J.\  {\bf 696} (2009) 608
  [arXiv:0809.4703 [astro-ph]].

\bibitem{Viaux:2013lha}
  N.~Viaux, M.~Catelan, P.~B.~Stetson, G.~Raffelt, J.~Redondo, A.~A.~R.~Valcarce and A.~Weiss,
  Phys.\ Rev.\ Lett.\  {\bf 111} (2013) 231301
  [arXiv:1311.1669 [astro-ph.SR]].


\bibitem{Georgi:1981pg}
  H.~M.~Georgi, S.~L.~Glashow and S.~Nussinov,
  Nucl.\ Phys.\ B {\bf 193} (1981) 297.
  
\bibitem{Grifols:1986fc}
  J.~A.~Grifols and E.~Masso,
  Phys.\ Lett.\ B {\bf 173} (1986) 237.

\bibitem{Hannestad:2007ys}
  S.~Hannestad, G.~Raffelt and Y.~Y.~Y.~Wong,
  Phys.\ Rev.\ D {\bf 76} (2007) 121701
  [arXiv:0708.1404 [hep-ph]].

\bibitem{Friedland:2007yj}
  A.~Friedland and M.~Giannotti,
  Phys.\ Rev.\ Lett.\  {\bf 100} (2008) 031602
  [arXiv:0709.2164 [hep-ph]].

\bibitem{Raffelt:1985nk} 
  G.~G.~Raffelt,
  Phys.\ Rev.\ D {\bf 33}, 897 (1986). 

\bibitem{Raffelt:1987yu}
  G.~G.~Raffelt and D.~S.~P.~Dearborn,
  Phys.\ Rev.\ D {\bf 36} (1987) 2211.

\bibitem{Friedland:2012hj}
  A.~Friedland, M.~Giannotti and M.~Wise,
  Phys.\ Rev.\ Lett.\  {\bf 110} (2013) 061101
  [arXiv:1210.1271 [hep-ph]].


\bibitem{Ayala:2014pea}
  A.~Ayala, I.~Dominguez, M.~Giannotti, A.~Mirizzi and O.~Straniero,
  arXiv:1406.6053 [astro-ph.SR].
  
\bibitem{Baker:2013zta}
  K.~Baker, G.~Cantatore, S.~A.~Cetin, M.~Davenport, K.~Desch, B.~Döbrich, H.~Gies and I.~G.~Irastorza {\it et al.},
  Annalen Phys.\  {\bf 525} (2013) A93
  [arXiv:1306.2841 [hep-ph]].

\bibitem{Rubakov:1997vp}
  V.~A.~Rubakov,
  JETP Lett.\  {\bf 65} (1997) 621
  [hep-ph/9703409].
  
\bibitem{Berezhiani:2000gh}
  Z.~Berezhiani, L.~Gianfagna and M.~Giannotti,
  Phys.\ Lett.\ B {\bf 500} (2001) 286
  [hep-ph/0009290].

\bibitem{Gianfagna:2004je}
  L.~Gianfagna, M.~Giannotti and F.~Nesti,
  JHEP {\bf 0410} (2004) 044
  [hep-ph/0409185].

\bibitem{Carosi:2013rla}
  G.~Carosi, A.~Friedland, M.~Giannotti, M.~J.~Pivovaroff, J.~Ruz and J.~K.~Vogel,
  arXiv:1309.7035 [hep-ph].

 \bibitem{Andriamonje:2007ew} 
  S.~Andriamonje {\it et al.}  [CAST Collaboration],
  JCAP {\bf 0704}, 010 (2007)
  [hep-ex/0702006].
  
\bibitem{Irastorza:2011gs}
  I.~G.~Irastorza, F.~T.~Avignone, S.~Caspi, J.~M.~Carmona, T.~Dafni, M.~Davenport, A.~Dudarev and G.~Fanourakis {\it et al.},
  JCAP {\bf 1106} (2011) 013
  [arXiv:1103.5334 [hep-ex]].

\bibitem{Lauterborn:1971nva} 
  D.~Lauterborn, S.~Refsdal and A.~Weigert,
  Astron.\ Astrophys.\  {\bf 10}, 97 (1971).
  
\bibitem{Keller:2008yu} 
  S.~C.~Keller,
  arXiv:0801.1342 [astro-ph].
 
 \bibitem{Salaris:2004xd} 
   M.~Salaris, M.~Riello, S.~Cassisi and G.~Piotto,
   Astron.\ and Astrophys.\  {\bf 420}, 911 (2004)
   [astro-ph/0403600].
    
\bibitem{Izotov:2013waa} 
  Y.~I.~Izotov, G.~Stasinska and N.~G.~Guseva,
  Astron.\ \& Astroph.\ {\bf 558}, A57 (2013)
  [arXiv:1308.2100 [astro-ph.CO]].
  
\bibitem{Aver:2013wba} 
  E.~Aver, K.~A.~Olive, R.~L.~Porter and E.~D.~Skillman,
  JCAP {\bf 1311}, 017 (2013)
  [arXiv:1309.0047 [astro-ph.CO]].

\bibitem{Bahre:2013ywa} 
  R.~Bähre, B.~Döbrich, J.~Dreyling-Eschweiler, S.~Ghazaryan, R.~Hodajerdi, D.~Horns, F.~Januschek and E.~-A.~Knabbe {\it et al.},
  JINST {\bf 8}, T09001 (2013)
  [arXiv:1302.5647 [physics.ins-det]].
   




  
  

\end{thebibliography}
\end{document}